\documentclass[9pt,conference]{IEEEtran}
\IEEEoverridecommandlockouts

\usepackage{cite}
\usepackage{amsmath,amssymb,amsfonts}
\usepackage{graphicx}
\usepackage{textcomp}
\usepackage{xcolor}
\usepackage{algorithm}
\usepackage{algpseudocode}
\usepackage{amsthm}

\usepackage{balance}

\DeclareMathOperator*{\minimize}{minimize}
\DeclareMathOperator*{\maximize}{maximize}

\newtheorem{theorem}{Theorem}

\def\BibTeX{{\rm B\kern-.05em{\sc i\kern-.025em b}\kern-.08em
		T\kern-.1667em\lower.7ex\hbox{E}\kern-.125emX}}
\begin{document}
	\title{Non-Convex Recovery from Phaseless Low-Resolution Blind Deconvolution Measurements using Noisy Masked Patterns	}
	
	\author{\IEEEauthorblockN{Samuel Pinilla$^{\dag}$, Kumar Vijay Mishra$^{\ddag}$ and Brian M. Sadler$^{\ddag}$}
		\IEEEauthorblockA{$^{\dag}$Faculty of Information Technology and Communication Sciences, Tampere University\\ 
			$^{\ddag}$United States CCDC Army Research Laboratory, Adelphi, MD 20783 USA }}
	
	\maketitle
	
	\begin{abstract}
		This paper addresses recovery of a kernel $\boldsymbol{h}\in \mathbb{C}^{n}$ and a signal $\boldsymbol{x}\in \mathbb{C}^{n}$ from the low-resolution phaseless measurements of their noisy circular convolution $\boldsymbol{y} = \left \rvert \boldsymbol{F}_{lo}( \boldsymbol{x}\circledast \boldsymbol{h}) \right \rvert^{2} + \boldsymbol{\eta}$, where $\boldsymbol{F}_{lo}\in \mathbb{C}^{m\times n}$ stands for a partial discrete Fourier transform ($m<n$), $\boldsymbol{\eta}$ models the noise, and $\lvert \cdot \rvert$ is the element-wise absolute value function. This problem is severely ill-posed because both the kernel and signal are unknown and, in addition, the measurements are phaseless, leading to many $\boldsymbol{x}$-$\boldsymbol{h}$ pairs that correspond to the measurements. Therefore, to guarantee a stable recovery of $\boldsymbol{x}$ and $\boldsymbol{h}$ from $\boldsymbol{y}$, we assume that the kernel $\boldsymbol{h}$ and the signal $\boldsymbol{x}$ lie in known subspaces of dimensions $k$ and $s$, respectively, such that $m\gg k+s$. We solve this problem by proposing a \textit{bli}nd deconvolution algorithm for \textit{pha}seless \textit{su}per-resolution (BliPhaSu) to minimize a non-convex least-squares objective function. The method first estimates a low-resolution version of both signals through a spectral algorithm, which are then refined based upon a sequence of stochastic gradient iterations. We show that our BliPhaSu algorithm converges linearly to a pair of true signals on expectation under a proper initialization that is based on spectral method. Numerical results from experimental data demonstrate perfect recovery of both $\boldsymbol{h}$ and $\boldsymbol{x}$ using our method.
	\end{abstract}
	
	
	\section{Introduction}
	Phase retrieval is a common problem that arises in numerous applications where only intensity measurements are available, such as astronomical imaging \cite{fienup1987phase}, microscopy \cite{mayo2003x}, X-ray crystallography \cite{millane1990phase,pinilla2018coded,pinilla2018crys}, and diffractive optical imaging (DOI) \cite{xu2015overcoming}. The interest in phase retrieval is largely because the imaging device is unable to measure the phase of optical signal. The problem is ill-posed for one-dimensional (1D) signals, implying that more than one signal with different phase has the same magnitude. Broadly, this problem is solved through either exploiting prior knowledge of signal structure such as sparse support \cite{vaswani2020nonconvex} or obtain additional measurements of the magnitude using, for example, masks \cite{katkovnik2017computational}. One of the main phase retrieval limitations in optical setups is the spatial resolution, which is limited by the optics and the sensor resolution \cite{yang2010image,park2003super}. This \textit{super-resolved phase retrieval}, i.e. estimation of high-resolution signals from low-resolution phaseless measurements, has been previously studied for coded diffraction patterns \cite{katkovnik2017computational,bacca2019super}, noiseless masks \cite{jaganathan2016phaseless}, or noiseless Fourier measurements \cite{chen2014algorithm}. 
	
	In particular, coded diffraction patterns modify the traditional DOI system by modulating the object with a mask and then collecting the intensity of its diffraction pattern \cite{candes2015phase,pinilla2018coded}. Further, the DOI involves the propagation of light through a medium where its rays are convolved with an unknown kernel. In applications such as visible light communications, the propagation of information carrying light through an unknown communications medium is modeled as a convolution. Since the channel is unknown, it is generally difficult at the receiver to obtain the phase information of the propagated light. This presents a challenging combination of super-resolution \cite{mishra2015spectral,xu2014precise,cho2015block}, phase retrieval \cite{fienup1987phase}, and blind deconvolution \cite{lee2018fast} problems in a single measurement system. Each one of these problems is ill-posed and, as a result, the combined problem becomes severely ill-posed. In this paper, we focus on the super-resolution phase retrieval involving blind deconvolution. This problem has remained unexamined in the previous works. 
	
	In particular, we investigate recovering a kernel and a signal from the low-resolution phaseless measurements of their noisy circular convolution. To guarantee a stable recovery, we assume that the kernel and the signal belong to known subspaces. We propose a \textit{bli}nd deconvolution algorithm for \textit{pha}seless \textit{su}per-resolution (BliPhaSu), which minimizes a nonconvex least-squares objective function. The method employs a spectral algorithm to estimate a low-resolution version of both kernel and signal, which is then refined based upon a sequence of stochastic gradient iterations. When a proper initialization based on spectral method is employed, our algorithm converges linearly to the pair of the true signals on expectation. We provide theoretical guarantees to analytically characterize the performance of the proposed initialization and the stochastic refining procedure. Numerical results from experimental data demonstrate that our BliPhaSu algorithm recovers both signals from noiseless and sparse noisy samples.
	
	The rest of the paper is organized as follows. In the next section, we discuss the system model and formulate the non-convex optimization problem to recover both kernel and signal. In Section \ref{sec:algorithm}, we provide a mathematical description of the proposed estimation procedure of the signals through the extraction of the leading eigenvectors of two designed matrices depending on the phaseless measurements. In Section \ref{sec:results}, we validate our models and methods through numerical experiments and experimental data. We conclude in Section \ref{sec:conclusion}.
	
	Throughout this paper, we use boldface lowercase  and uppercase letters for vectors and matrices, respectively. We denote by $\mathbb{R}_{+}:=\{w\in \mathbb{R}: w\geq 0\}$ and $\mathbb{R}_{++}:=\{w\in \mathbb{R}: w>0\}$ the sets of positive and strictly positive real numbers, respectively. The conjugate and the conjugate transpose of the vector $\boldsymbol{w}\in \mathbb{C}^{n}$ are denoted as $\overline{\boldsymbol{w}}\in \mathbb{C}^{n}$ and $\boldsymbol{w}^{H}\in \mathbb{C}^{n}$, respectively. The notation $\circledast$ denotes circular convolution operation. The $\ell$th entry of a vector $\boldsymbol{w}$, is $\boldsymbol{w}[\ell]$. For a matrix, the $(a,b)$ entry of $\boldsymbol{W}\in \mathbb{C}^{m\times n}$ is denoted by $\boldsymbol{W}[a,b]$.
	
	\section{Problem Formulation}
	\label{sec:problem}
	Assume $\boldsymbol{h},\boldsymbol{x}\in \mathbb{C}^{n}$ to be vectors, and define the low-frequency discrete Fourier transform (DFT) matrix $\boldsymbol{F}_{lo}\in \mathbb{R}^{m\times n}$ with $m<n$. Then, we observe the noisy low-resolution phaseless measurements of the circular convolution between $\boldsymbol{h}$ and $\boldsymbol{x}$ as 
	\begin{align}
	\boldsymbol{y} = \left \rvert \boldsymbol{F}_{lo}(\boldsymbol{x}\circledast \boldsymbol{h}) \right \rvert^{2} + \boldsymbol{\eta},
	\label{eq:basicproblem}
	\end{align}
	where $\boldsymbol{\eta}$ models the noise, and $\lvert \cdot \rvert$ is the element-wise absolute value function. To analytically address the problem, we assume that both $\boldsymbol{h}$ and $\boldsymbol{x}$ are members of known subspaces of $\mathbb{C}^{n}$. This means that $\boldsymbol{h}$ and $\boldsymbol{x}$ can be parameterized in terms of unknown lower dimensional vectors $\boldsymbol{g}\in \mathbb{C}^{k}$ and $\boldsymbol{z}\in \mathbb{C}^{s}$, respectively, as follows $\boldsymbol{h} = \boldsymbol{B}\boldsymbol{g}, \text{  }\boldsymbol{x} = \boldsymbol{C}\boldsymbol{z},$ where $\boldsymbol{B}\in \mathbb{C}^{n\times k}$, and $\boldsymbol{C}\in \mathbb{C}^{n\times s}$ are known matrices whose columns span the subspaces in which $\boldsymbol{h}$ and $\boldsymbol{x}$ belong, respectively. 
	
	The circular convolution operator diagonalizes in the Fourier domain. Therefore, \eqref{eq:basicproblem} becomes \cite{ahmed2019simultaneous}
	\begin{align}
	\boldsymbol{y} = \left \rvert \hat{\boldsymbol{B}}\boldsymbol{g}\odot \hat{\boldsymbol{C}}\boldsymbol{z} \right \rvert^{2} + \boldsymbol{\eta},
	\label{eq:newproblem}
	\end{align}
	with $\hat{\boldsymbol{B}} = \boldsymbol{F}_{lo}\boldsymbol{B}\in \mathbb{C}^{m\times k}$, $\hat{\boldsymbol{C}} = \boldsymbol{F}_{lo}\boldsymbol{C}\in \mathbb{C}^{m\times s}$, $\odot$ represents the Hadamard product, and $\boldsymbol{F}$ stands for the normalized DFT matrix. Therefore, from \eqref{eq:newproblem}, we are interested in recovering $\boldsymbol{g}$ and $\boldsymbol{z}$ from the phaseless measurements $\boldsymbol{y}$. Note that \eqref{eq:newproblem} is a combination of three problems: super-resolution ($\boldsymbol{F}_{lo}$ represents low-resolution measurements), blind deconvolution (both $\boldsymbol{g}$ and $\boldsymbol{z}$ are unknown), and phase retrieval ($\boldsymbol{y}$ is phaseless).
	
	Our goal is to estimate both $\boldsymbol{g}$ and $\boldsymbol{z}$ from the phaseless data $\boldsymbol{y}$. To this end, we propose a non-convex optimization problem. Observe that the $\ell$-th entry of vector $\boldsymbol{y}$ from \eqref{eq:newproblem} is 
	\begin{align}
	\boldsymbol{y}[\ell] = \left \lvert (\hat{\boldsymbol{b}}_{\ell}^{H}\boldsymbol{g})(\hat{\boldsymbol{c}}_{\ell}^{H}\boldsymbol{z}) \right\rvert^{2} + \boldsymbol{\eta}[\ell],
	\label{eq:constraint}
	\end{align}
	for $\ell=1,\dots,m$, where $\hat{\boldsymbol{b}}_{\ell}^{H}$, and $\hat{\boldsymbol{c}}_{\ell}^{H}$ are the $\ell$-th rows of the matrices $\hat{\boldsymbol{B}}$ and $\hat{\boldsymbol{C}}$, respectively. Note that, in \eqref{eq:newproblem}, the matrices $\boldsymbol{B}$ and $\boldsymbol{C}$ are known and they parameterize the subspaces in which $\boldsymbol{h}$ and $\boldsymbol{x}$, respectively, lie. Then, to control the nature of these subspaces, consider $\boldsymbol{B}$, and $\boldsymbol{C}$ to follow a complex Gaussian distribution such that their columns satisfy $\boldsymbol{b}_{\ell},\boldsymbol{c}_{\ell}\sim \mathcal{CN}(\boldsymbol{0},\sigma^{2}\boldsymbol{I})$ (with $\sigma$ as the standard deviation). In consequence, matrices $\hat{\boldsymbol{B}}=\boldsymbol{F}_{lo}\boldsymbol{B}$ and $\hat{\boldsymbol{C}}= \boldsymbol{F}_{lo}\boldsymbol{B}$ follow a Gaussian distribution because $\boldsymbol{F}_{lo}$ is a partial version of the DFT matrix. Without loss of generality, we choose the Gaussian distribution of $\boldsymbol{b}_{\ell}$ and $\boldsymbol{c}_{\ell}$ such that $\hat{\boldsymbol{b}}_{\ell},\hat{\boldsymbol{c}}_{\ell}\sim \mathcal{CN}(\boldsymbol{0},\boldsymbol{I})$. 
	
	In recent years, substantial work has been done and is still ongoing to recover a signal from phaseless quadratic random measurements. A popular approach is to minimize the intensity least-squares objective; see for instance \cite{pinilla2018phase,pinilla2019frequency}. Similar to this technique, in order to estimate $\boldsymbol{g}$, and $\boldsymbol{z}$, we solve the following optimization problem
	\begin{align}
	\minimize_{\boldsymbol{g}\in \mathbb{C}^{k},\boldsymbol{z}\in \mathbb{C}^{s}} \hspace{1em}f(\boldsymbol{g},\boldsymbol{z})=\frac{1}{2m}\sum_{\ell=1}^{m}\left(\boldsymbol{y}[\ell] - \left \lvert (\hat{\boldsymbol{b}}_{\ell}^{H}\boldsymbol{g})(\hat{\boldsymbol{c}}_{\ell}^{H}\boldsymbol{z}) \right\rvert^{2} \right)^{2}.
	\label{eq:optimizing}
	\end{align} 
	The key idea 
	is to find a tuple $(\boldsymbol{g}_{*},\boldsymbol{z}_{*})$ that is most aligned with $(\boldsymbol{g},\boldsymbol{z})$ and satisfies a the measurement constraints in \eqref{eq:constraint}. Thus, equivalently, the signals $\boldsymbol{x}$, and $\boldsymbol{h}$ are approximated as $\hat{\boldsymbol{B}}\boldsymbol{g}_{*}$ and $\hat{\boldsymbol{C}}\boldsymbol{z}_{*}$, respectively. An important feature of this optimization problem is that it works in the same dimension of the signals and a lifting version of the problem is no longer needed, as in the previous works on phase retrieval~\cite{ahmed2018blind,candes2013phaselift,waldspurger2015phase}. 
	
	Note that the vectors $\mathbf{w}$ and $\mathbf{w}e^{j\theta}$ yield the same magnitude measurements for any constant global phase $\theta \in \mathbb{R}$, which is not recoverable. This ambiguity from global phase leads to the following performance metric between vectors $\boldsymbol{w}_{1}$ and $\boldsymbol{w}_{2}$: 
	\begin{align}
	\textup{ relative error} := \frac{\textrm{dist}(\boldsymbol{w}_{1},\boldsymbol{w}_{2})}{\lVert \boldsymbol{w}_{2} \rVert_{2}}, 
	\label{eq:relError}
	\end{align}
	where $\displaystyle \textrm{dist}(\boldsymbol{w}_{1},\boldsymbol{w}_{2}):= \minimize_{\theta \in [0,2\pi)} \lVert\boldsymbol{w}_{1}e^{-j\theta} - \boldsymbol{w}_{2}\rVert_{2}$ is the Euclidean distance modulo a global unimodular constant between two complex vectors. If $\textrm{dist}\left(\boldsymbol{w}_1,\boldsymbol{w}_2\right)=0$, then $\boldsymbol{w}_1$ and $\boldsymbol{w}_2$ are equal up to some global phase.
	
	\section{Non-convex Reconstruction Algorithm}
	\label{sec:algorithm}
	To solve \eqref{eq:optimizing}, we propose a stochastic gradient algorithm that is initialized by a spectral procedure to estimate $\boldsymbol{g}$ and $\boldsymbol{z}$. 
	
	\subsection{Initialization Step}
	Assume a noise-free measurements vector $\boldsymbol{y}$ and define the auxiliary matrix 
	\begin{align}
	\boldsymbol{H}^{\ell}_{g} &= \boldsymbol{y}[\ell] \hat{\boldsymbol{b}}_{\ell}\hat{\boldsymbol{b}}_{\ell}^{H} \nonumber\\
	&= \left\lvert \sum_{p=1}^{k}\sum_{q=1}^{s} \overline{\hat{\boldsymbol{b}}}_{\ell}[p]\boldsymbol{g}[p] \overline{\hat{\boldsymbol{c}}}_{\ell}[q]\boldsymbol{z}[q] \right\rvert^{2}\hat{\boldsymbol{b}}_{\ell}\hat{\boldsymbol{b}}_{\ell}^{H} \nonumber\\
	&= \sum_{p_{i}=1}^{k}\sum_{q_{i}=1}^{s} \boldsymbol{g}[p_{1}]\overline{\boldsymbol{g}}[p_{2}] \boldsymbol{z}[q_{1}]\overline{\boldsymbol{z}}[q_{2}] \overline{\hat{\boldsymbol{b}}}_{\ell}[p_{1}]\hat{\boldsymbol{b}}_{\ell}[p_{2}] \overline{\hat{\boldsymbol{c}}}_{\ell}[q_{1}]\hat{\boldsymbol{c}}_{\ell}[q_{2}] \hat{\boldsymbol{b}}_{\ell}\hat{\boldsymbol{b}}_{\ell}^{H},
	\end{align}
	for $i=1,2$. Recall that vectors $\hat{\boldsymbol{b}}_{\ell}^{H}$ and $\hat{\boldsymbol{c}}_{\ell}^{H}$ follow a Gaussian distribution with zero-mean and unit variance. This implies that the expected value of the main diagonal is 
	\begin{equation}
	\mathbb{E}\left[\boldsymbol{H}^{\ell}_{g}[r,r]\right] = \left\lbrace \begin{array}{ll}
	\lVert \boldsymbol{g} \rVert^{2}_{2} \lVert \boldsymbol{z} \rVert^{2}_{2} & p_{1}=p_{2}, q_{1}=q_{2} \\
	0 & \text{ otherwise }
	\end{array}\right..
	\label{eq:firstpartg}
	\end{equation}
	Additionally, for those off-diagonal entries of matrix $\boldsymbol{H}^{\ell}_{g}$, we have
	\begin{equation}
	\mathbb{E}\left[\boldsymbol{H}^{\ell}_{g}[r,a]\right] = \left\lbrace\begin{array}{ll}
	\lVert \boldsymbol{z} \rVert_{2}^{2}(\boldsymbol{g}[r]\overline{\boldsymbol{g}}[a]) & p_{1}=r,p_{2}=a, q_{1}=q_{2} \\
	0 & \text{ otherwise }
	\end{array}\right..
	\label{eq:secondpartg}
	\end{equation}
	Combining \eqref{eq:firstpartg} and \eqref{eq:secondpartg}, we obtain
	\begin{align}
	\mathbb{E}\left[\boldsymbol{H}^{\ell}_{g}\right] = \lVert \boldsymbol{z}\rVert_{2}^{2}\left(\lVert \boldsymbol{g} \rVert^{2}_{2}\boldsymbol{I} + \boldsymbol{g}\boldsymbol{g}^{H}\right).
	\end{align}
	
	Following the strong law of large numbers, the sample average approaches the ensemble one namely, as $m$ increases,  $\displaystyle\boldsymbol{H}_{g} = \frac{1}{m}\sum_{\ell=1}^{m} \boldsymbol{H}^{\ell}_{g} \rightarrow \mathbb{E}\left[ \boldsymbol{H}^{\ell}_{g}\right]$. Thus, considering that the largest eigenvector of $\boldsymbol{H}_{g}$ is aligned with $\boldsymbol{g}$, we approximate $\boldsymbol{g}$ by solving
	\begin{align}
	\maximize_{\lVert \boldsymbol{w} \rVert_{2}=1} \hspace{0.5em}\boldsymbol{w}^{H}\left(\frac{1}{m}\sum_{\ell=1}^{m} \boldsymbol{y}[\ell] \hat{\boldsymbol{b}}_{\ell}\hat{\boldsymbol{b}}_{\ell}^{H}\right)\boldsymbol{w}.
	\label{eq:solg}
	\end{align}
	Then, the low-dimensional representation vector of the kernel $\boldsymbol{h}$ in $\hat{\boldsymbol{B}}$ is $\boldsymbol{g}$. Assume, without loss of generality, $\lVert \boldsymbol{g}\rVert_{2}=1$. Thus, the initial estimation of $\boldsymbol{g}$ is defined as $\boldsymbol{g}^{(0)}=\boldsymbol{w}_{g}$, where $\boldsymbol{w}_{g}$ is the solution vector of \eqref{eq:solg}. As a consequence, the initial approximation $\boldsymbol{h}^{(0)}$ of the kernel $\boldsymbol{h}$ is  $\boldsymbol{h}^{(0)} = \boldsymbol{B}\boldsymbol{g}^{(0)}$.
	
	Proceeding with the vector $\boldsymbol{z}$, define the matrix
	\begin{align}
	\boldsymbol{H}^{\ell}_{z} &= \boldsymbol{y}[\ell] \hat{\boldsymbol{c}}_{\ell}\hat{\boldsymbol{c}}_{\ell}^{H} \nonumber\\
	&= \sum_{p_{i}=1}^{k}\sum_{q_{i}=1}^{s} \boldsymbol{g}[p_{1}]\overline{\boldsymbol{g}}[p_{2}] \boldsymbol{z}[q_{1}]\overline{\boldsymbol{z}}[q_{2}] \overline{\hat{\boldsymbol{b}}}_{\ell}[p_{1}]\hat{\boldsymbol{b}}_{\ell}[p_{2}] \overline{\hat{\boldsymbol{c}}}_{\ell}[q_{1}]\hat{\boldsymbol{c}}_{\ell}[q_{2}] \hat{\boldsymbol{c}}_{\ell}\hat{\boldsymbol{c}}_{\ell}^{H}.
	\label{eq:initz}
	\end{align}
	Repeating the process over \eqref{eq:initz} as in \eqref{eq:firstpartg} and \eqref{eq:secondpartg} yields
	\begin{align}
	\mathbb{E}\left[\boldsymbol{H}^{\ell}_{z} \right] &= \lVert \boldsymbol{g}\rVert_{2}^{2}\left(\lVert \boldsymbol{z} \rVert^{2}_{2}\boldsymbol{I} + \boldsymbol{z}\boldsymbol{z}^{H}\right) \nonumber\\
	&=\lVert \boldsymbol{z} \rVert^{2}_{2}\boldsymbol{I} + \boldsymbol{z}\boldsymbol{z}^{H},
	\end{align}
	where the second equality comes from the assumption that $\lVert \boldsymbol{g}\rVert_{2}=1$. Thus, following the strong law of large numbers, the sample average approaches the ensemble one: $\displaystyle\boldsymbol{H}_{z} = \lim_{m\rightarrow \infty} \frac{1}{m}\sum_{\ell=1}^{m} \boldsymbol{H}^{\ell}_{z} \rightarrow \mathbb{E}\left[ \boldsymbol{H}^{\ell}_{z}\right]$. Then, the largest eigenvector of $\boldsymbol{H}_{z}$ being aligned with $\boldsymbol{z}$, we approximate $\boldsymbol{z}$ by solving the problem
	\begin{align}
	\maximize_{\lVert \boldsymbol{w} \rVert_{2}=1} \hspace{0.5em}\boldsymbol{w}^{H}\left(\frac{1}{m}\sum_{\ell=1}^{m} \boldsymbol{y}[\ell] \hat{\boldsymbol{c}}_{\ell}\hat{\boldsymbol{c}}_{\ell}^{H}\right)\boldsymbol{w}.
	\label{eq:solz}
	\end{align}
	Therefore, taking $\boldsymbol{w}_{z}$ as the solution vector of \eqref{eq:solz} and $\lambda_{z}$ as the leading eigenvalue of matrix $\boldsymbol{H}_{z}$, the initial estimation of $\boldsymbol{z}$ is  $\boldsymbol{z}^{(0)}=\sqrt{\frac{\lambda_{z}}{2}}\boldsymbol{w}_{z}$. Additionally, the initial approximation $\boldsymbol{x}^{(0)}$ of the signal $\boldsymbol{x}$ is $\boldsymbol{x}^{(0)} = \boldsymbol{C}\boldsymbol{z}^{(0)}$.
	
	The following Theorem~\ref{theo:initialization} analytically characterizes the closeness of the initial estimations $\boldsymbol{g}^{(0)}$ and $\boldsymbol{z}^{(0)}$ with $\boldsymbol{g}$ and $\boldsymbol{z}$ respectively.
	\begin{theorem}
		Consider $\boldsymbol{y}$ in \eqref{eq:newproblem} such that $\lVert \boldsymbol{\eta} \rVert_{\infty}\leq \rho\lVert \boldsymbol{z} \rVert_{2}$. Assume $\boldsymbol{z}^{(0)} = \sqrt{\frac{\lambda_{z}}{2}}\boldsymbol{w}_{z}$, and $\boldsymbol{g}^{(0)}=\boldsymbol{w}_{g}$ where $\boldsymbol{w}_{g},\boldsymbol{w}_{z}$ are the solutions of \eqref{eq:solg}, \eqref{eq:initz} respectively. Then,
		\begin{align}
		\text{ dist }(\boldsymbol{g}^{(0)},\boldsymbol{g}) &< \tau_{g} + \mathcal{O}(\lVert \boldsymbol{\eta} \rVert_{\infty}), \label{eq:initialization}\\
		\text{ dist }(\boldsymbol{z}^{(0)},\boldsymbol{z}) &< \tau_{z} \lVert \boldsymbol{z} \rVert_{2} + \mathcal{O}(\lVert \boldsymbol{\eta} \rVert_{\infty}),
		\label{eq:initialization1}
		\end{align}
		for some $\tau_{g},\tau_{z}\in (0,1)$, and $m\geq \beta (k+s)$ with $\beta>0$ sufficiently large constant.
		\label{theo:initialization}
	\end{theorem}
	\begin{proof}
		See \cite{Senn:2009}.
	\end{proof}
	Observe that Theorem \ref{theo:initialization} essentially guarantees that $\boldsymbol{g}^{(0)}$ ($\boldsymbol{z}^{(0)}$) is an acceptable initial estimation of $\boldsymbol{g}$ ($\boldsymbol{z}$). This initial step needs to be refined, as explained next. 
	
	\subsection{Stochastic Gradient Refinement Step}
	Here, we use the theory of Wirtinger derivatives \cite{hunger2007introduction}. The gradient of $f(\boldsymbol{g},\boldsymbol{z})$ in \eqref{eq:optimizing} with respect to $\boldsymbol{g}$ is 
	\begin{align}
	\nabla_{\boldsymbol{g}}f(\boldsymbol{g},\boldsymbol{z})  := \left[\frac{\partial f(\boldsymbol{g},\boldsymbol{z})}{\partial \overline{\boldsymbol{g}}[0]},\cdots,\frac{\partial f(\boldsymbol{g},\boldsymbol{z})}{\partial \overline{\boldsymbol{g}}[N-1]}\right]^{H}.
	\label{eq:gradient}
	\end{align}
	The definition of $\nabla_{\boldsymbol{z}}f(\boldsymbol{g},\boldsymbol{z})$ is analogously derived from \eqref{eq:gradient}. 
	Define a standard gradient algorithm for $\boldsymbol{g}$ as
	\begin{align}
	\boldsymbol{g}^{(t+1)}:=\boldsymbol{g}^{(t)} - \alpha_{g} \nabla_{\boldsymbol{g}}f(\boldsymbol{g}^{(t)},\boldsymbol{z}^{(t)}),
	\label{eq:updateforx}
	\end{align}
	where $\alpha_{g}$ is the step size. The definition of the standard gradient step for $\nabla_{\boldsymbol{z}}f(\boldsymbol{g},\boldsymbol{z})$ is analogously derived from \eqref{eq:updateforx} with step size $\alpha_{z}$. 
	
	To alleviate the memory requirements and computational complexity for large $m$, we suggest a stochastic gradient descent strategy. Instead of computing \eqref{eq:updateforx}, we choose only a random subset of the sum for each iteration $t$ leading to following refinement steps
	\begin{equation}
	\boldsymbol{g}^{(t+1)}:=\boldsymbol{g}^{(t)} -  \frac{\alpha_{g}}{m}\sum_{\ell\in \Gamma_{(t)}} \boldsymbol{\gamma}_{t}[\ell] \left \lvert \hat{\boldsymbol{c}}_{\ell}^{H}\boldsymbol{z}^{(t)} \right\rvert^{2} \hat{\boldsymbol{b}}_{\ell}\hat{\boldsymbol{b}}_{\ell}^{H}\boldsymbol{g}^{(t)},
	\label{eq:functiong}
	\end{equation}
	where
	\begin{align}
	\boldsymbol{z}^{(t+1)}:=\boldsymbol{z}^{(t)} -  \frac{\alpha_{z}}{m}\sum_{\ell\in \Gamma_{(t)}} \boldsymbol{\gamma}_{t}[\ell] \left \lvert \hat{\boldsymbol{b}}_{\ell}^{H}\boldsymbol{g}^{(t)} \right\rvert^{2} \hat{\boldsymbol{c}}_{\ell}\hat{\boldsymbol{c}}_{\ell}^{H}\boldsymbol{z}^{(t)},
	\label{eq:functionz}
	\end{align}
	and $\boldsymbol{\gamma}_{t}[\ell]= \left \lvert (\hat{\boldsymbol{b}}_{\ell}^{H}\boldsymbol{g}^{(t)})(\hat{\boldsymbol{c}}_{\ell}^{H}\boldsymbol{z}^{(t)}) \right\rvert^{2} - \boldsymbol{y}[\ell]$. In \eqref{eq:functiong} and \eqref{eq:functionz}, the set $\Gamma_{(t)}$ is chosen uniformly and independently at random at each iteration $t$ from subsets of $\{1,\cdots,m \}$ with cardinality $Q$. Specifically, the gradient in \eqref{eq:gradient} is uniformly sampled using a minibatch of data (in this case, of size $Q$ for each update) such that, in expectation, it is $\nabla_{\boldsymbol{g}}f(\boldsymbol{g},\boldsymbol{z})$ (analogously defined for $\nabla_{\boldsymbol{z}}f(\boldsymbol{g},\boldsymbol{z})$) \cite[page 130]{spall2005introduction}.
	
	\begin{algorithm}[t!]
		\caption{\textit{Bli}nd deconvolution for \textit{pha}seless \textit{su}per-resolution (BliPhaSu)}
		\label{alg:algorithm}
		\small
		\begin{algorithmic}[1]
			\Statex{\textbf{Input: }Data $\left\lbrace\boldsymbol{y}[\ell]:\ell=1,\cdots,m \right\rbrace$. Choose the constants $\alpha_{g},\alpha_{z}>0$, $tol=1\times 10^{-2}$, the integer constant $Q$, and matrices $\boldsymbol{B},\boldsymbol{C}, \hat{\boldsymbol{B}}$, and $\hat{\boldsymbol{C}}$.}
			\Statex{\textbf{Output:} $\boldsymbol{x}^{(T)}, \boldsymbol{h}^{(T)}$}
			\State{Compute $\boldsymbol{H}_{g}$, and $\boldsymbol{H}_{z}$
				\begin{align*}
				\boldsymbol{H}_{g} = \frac{1}{m}\sum_{\ell=1}^{m} \boldsymbol{y}[\ell] \hat{\boldsymbol{b}}_{\ell}\hat{\boldsymbol{b}}_{\ell}^{H}, \hspace{1em}\boldsymbol{H}_{z} & = \frac{1}{m}\sum_{\ell=1}^{m} \boldsymbol{y}[\ell] \hat{\boldsymbol{c}}_{\ell}\hat{\boldsymbol{c}}_{\ell}^{H}
				\end{align*}}
			\State {Extract the leading eigenvector $\boldsymbol{w}_{g}$, and $\boldsymbol{w}_{g}$ of the matrices $\boldsymbol{H}_{g}$, and $\boldsymbol{H}_{z}$, respectively. Define,
				\begin{align*}
				\boldsymbol{z}^{(0)} = \sqrt{\frac{\lambda_{z}}{2}}\boldsymbol{w}_{z}, \hspace{1em} \boldsymbol{g}^{(0)}=\boldsymbol{w}_{g}
				\end{align*}	
				with $\lambda_{z}$ as the leading eigenvalue of matrix $\boldsymbol{H}_{z}$.}
			\Statex{}
			\While{$\lVert \boldsymbol{d}_{g}^{(t)} \rVert_{2} \text{ and } \lVert \boldsymbol{d}_{z}^{(t)}  \rVert_{2}\geq tol$}
			\State{Choose $\Gamma_{(t)}$ uniformly and independently at random.}
			\State{Compute
				\begin{align*}
				\boldsymbol{g}^{(t+1)}:=\boldsymbol{g}^{(t)} -  \frac{\alpha_{g}}{m}\underbrace{\sum_{\ell\in \Gamma_{(t)}} \boldsymbol{\gamma}_{t}[\ell] \left \lvert \hat{\boldsymbol{c}}_{\ell}^{H}\boldsymbol{z}^{(t)} \right\rvert^{2} \hat{\boldsymbol{b}}_{\ell}\hat{\boldsymbol{b}}_{\ell}^{H}\boldsymbol{g}^{(t)}}_{\boldsymbol{d}_{g}^{(t)} }
				\end{align*}
			}
			\State{and
				\begin{align*}
				\boldsymbol{z}^{(t+1)}:=\boldsymbol{z}^{(t)} -  \frac{\alpha_{z}}{m}\underbrace{\sum_{\ell\in \Gamma_{(t)}} \boldsymbol{\gamma}_{t}[\ell] \left \lvert \hat{\boldsymbol{b}}_{\ell}^{H}\boldsymbol{g}^{(t)} \right\rvert^{2} \hat{\boldsymbol{c}}_{\ell}\hat{\boldsymbol{c}}_{\ell}^{H}\boldsymbol{z}^{(t)}}_{\boldsymbol{d}_{z}^{(t)} }
				\end{align*}
				with $\boldsymbol{\gamma}_{t}[\ell]= \left \lvert (\hat{\boldsymbol{b}}_{\ell}^{H}\boldsymbol{g}^{(t)})(\hat{\boldsymbol{c}}_{\ell}^{H}\boldsymbol{z}^{(t)}) \right\rvert^{2} - \boldsymbol{y}[\ell]$}
			\EndWhile
			\State{\textbf{return: } $\boldsymbol{x}^{(T)}=\boldsymbol{C}\boldsymbol{z}^{(T)}$, $\boldsymbol{h}^{(T)}=\boldsymbol{B}\boldsymbol{g}^{(T)}$. \Comment{$T$ indexes last iteration }}
		\end{algorithmic}
	\end{algorithm}
	
	Algorithm \ref{alg:algorithm} summarizes the BliPhaSu estimation steps for $\boldsymbol{g}$ and $\boldsymbol{z}$. The following Theorem~\ref{eq:convergence} characterizes the BliPhaSu convergence behavior. 
	\begin{theorem} 
		Consider $\boldsymbol{y}$ in \eqref{eq:newproblem} such that $\lVert \boldsymbol{\eta} \rVert_{\infty}\leq \rho\lVert \boldsymbol{z} \rVert_{2}$. The set $\Gamma_{(t)}$ is sampled uniformly at random from all subsets of $\{1,\cdots,m \}$, with cardinality $Q$, independently for each iteration. Then, BliPhaSu algorithm 
		with step sizes $\alpha_{g},\alpha_{z}\in (0,\frac{2}{U}] $ satisfies
		\begin{align}
		\text{ dist }(\boldsymbol{g}^{(t+1)},\boldsymbol{g}) &< \tau_{g}(1-\rho_{g})^{(t+1)} + \mathcal{O}(\lVert \boldsymbol{\eta} \rVert_{\infty}),\\
		\text{ dist }(\boldsymbol{z}^{(t+1)},\boldsymbol{z}) &< \tau_{z} (1-\rho_{z})^{(t+1)} \lVert \boldsymbol{z} \rVert_{2} + \mathcal{O}(\lVert \boldsymbol{\eta} \rVert_{\infty}),
		\label{eq:convergence}
		\end{align}
		for some constant $U>0$ depending on $\rho_{g},\rho_{z}\in (0,1)$.
		\label{theo:contraction}
	\end{theorem}
	\begin{proof}
		See \cite{Senn:2009}.
	\end{proof}
	
	\section{Numerical Results}
	\label{sec:results}
	We 
	validated our proposed approach using BliPhaSu algorithm
	through numerical experiments. 
	For all simulated experiments, the signals $\boldsymbol{x}$ and $\boldsymbol{h}$ are complex Gaussian vectors with $k=s=50$. We evaluate the performance with the empirical success rate among 100 trial runs. For each trial, $500$ iterations for all algorithms are employed\footnote{All simulations were performed using Matlab R2019a on an Intel Core i7 3.41Ghz CPU with 16 GB RAM.}. We declare that a trial is successful when the returned estimate incurs a relative error less than $10^{-5}$.
	
	We conducted three tests to evaluate the performance of the proposed method under noisy and noiseless scenarios at different values of signal-to-noise-ratio (SNR)$ = 10\log_{10}(\lVert \boldsymbol{y} \rVert^{2}_{\text{2}}/\lVert \boldsymbol{\sigma} \rVert^{2}_{\text{2}})$, where $\boldsymbol{\sigma}$ is the variance of the noise. First, we examine the performance of $\boldsymbol{g}^{(0)}$, and $\boldsymbol{z}^{(0)}$ to estimate $\boldsymbol{g}$, and $\boldsymbol{z}$, respectively. Then, we assess the performance of $\boldsymbol{g}^{(0)}$, $\boldsymbol{z}^{(0)}$ on the empirical success rate. Finally, we present an example of the estimated signal $\boldsymbol{x}$ and the kernel $\boldsymbol{h}$ from real experimental data.
	
	\subsection{Simulated Results}
	\begin{figure}[t]
		\centering
		\includegraphics[width=1\linewidth]{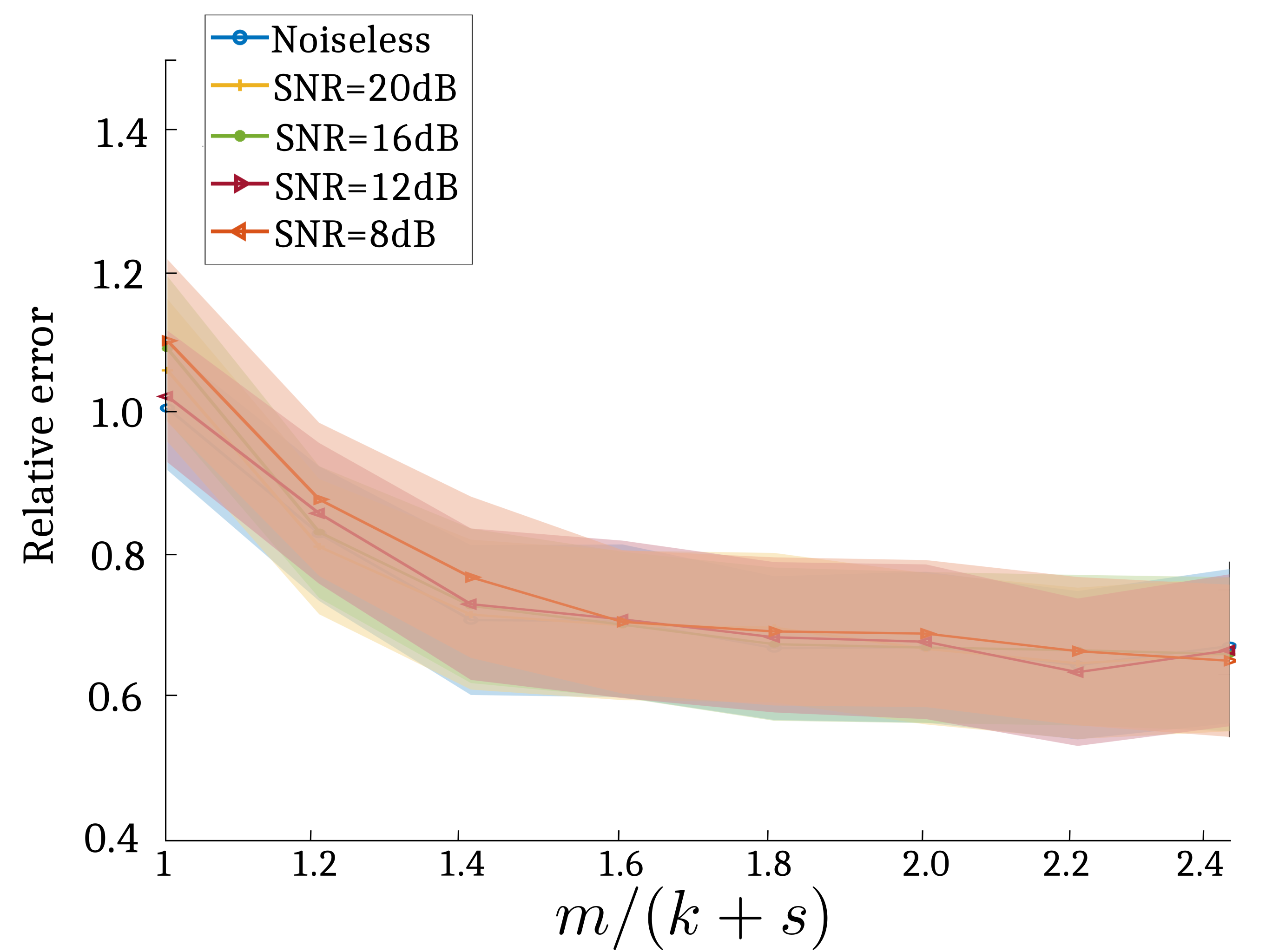}
		\caption{\footnotesize{Performance of $\boldsymbol{g}^{(0)}$ and $\boldsymbol{z}^{(0)}$ obtained by solving \eqref{eq:solg} and \eqref{eq:initz}, respectively, through a power iteration strategy at different SNR levels over $\boldsymbol{y}$ in \eqref{eq:newproblem} for different ratios of $m/(k+s)$. The relative error was averaged over 100 trials.}}
		\label{fig:init}		
	\end{figure}
	\begin{figure}[t]
		\centering
		\includegraphics[width=1.0\linewidth]{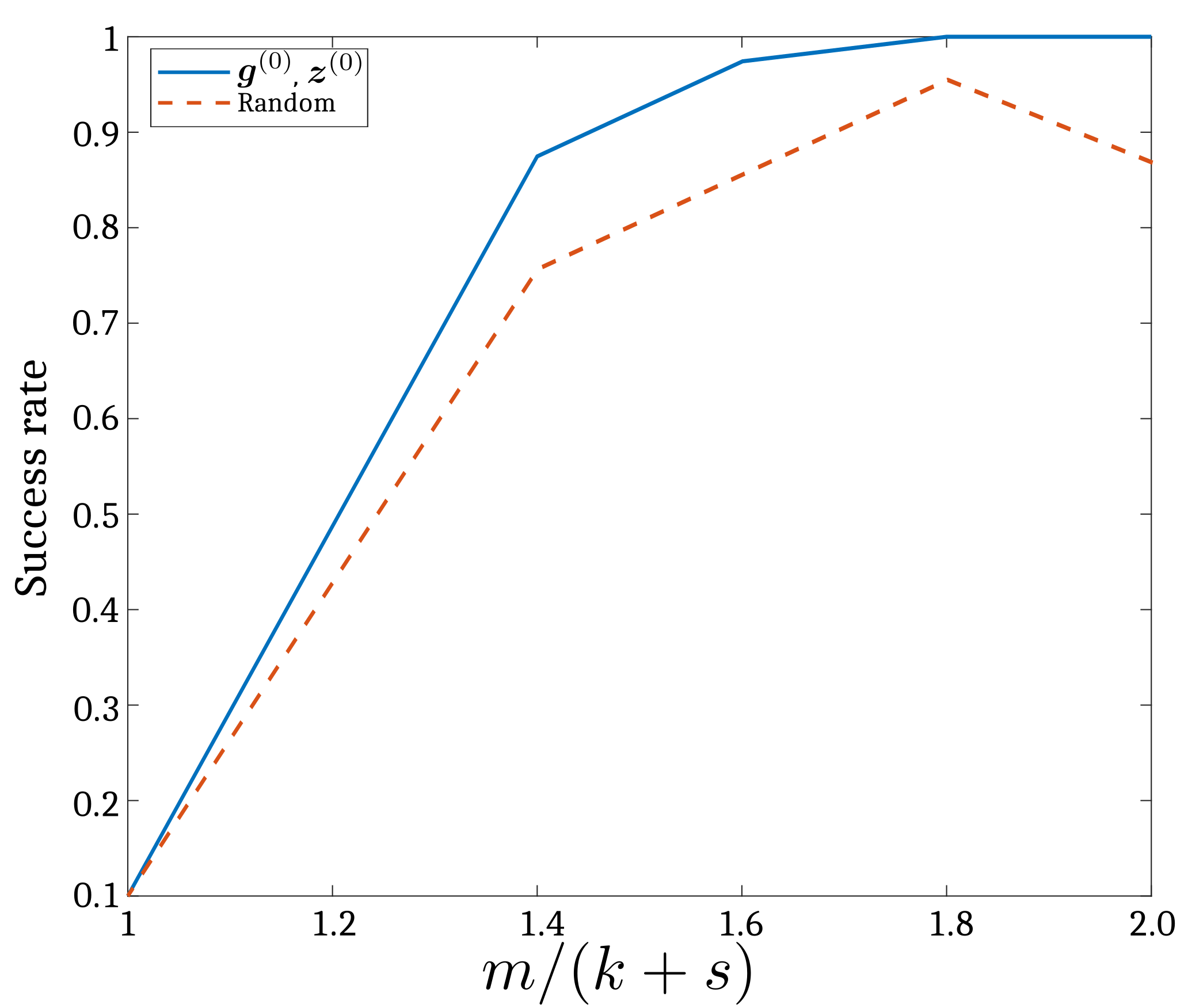}
		\caption{\footnotesize{Empirical success rate of BliPhaSu to solve \eqref{eq:optimizing} in the absence of noise. The dashed line is when both $\boldsymbol{g}^{(0)}$, $\boldsymbol{z}^{(0)}$ are random vectors. The solid lines is the returned solution of \eqref{eq:solg}, and \eqref{eq:initz}.}}
		\label{fig:success}
	\end{figure}
	We examined the performance of the returned initial points $\boldsymbol{g}^{(0)}$ and $\boldsymbol{z}^{(0)}$ obtained by solving \eqref{eq:solg} and \eqref{eq:initz}, respectively, under noisy (i.e., when $\boldsymbol{\eta}$ in \eqref{eq:newproblem} is white noise) and noiseless scenarios. Since both \eqref{eq:solg} and \eqref{eq:initz} involve the computation of the leading eigenvector of a matrix, we follow a power iteration strategy to estimate them. The number of iterations of this method was set to $150$. We numerically determined the average relative error $\left(\frac{\text{ dist }(\boldsymbol{g}^{(0)},\boldsymbol{g})}{2\lVert \boldsymbol{g}\rVert_{2}} + \frac{\text{ dist }(\boldsymbol{z}^{(0)},\boldsymbol{z})}{2\lVert \boldsymbol{z}\rVert_{2}}\right)$ as in \eqref{eq:relError}, averaged over 100 trials (Fig.~\ref{fig:init}). The results suggest the effectiveness of solving \eqref{eq:solg} and \eqref{eq:initz} to estimate the underlying signals.
	
	To complement the results in Fig. \ref{fig:init}, we studied the empirical success rate of solving \eqref{eq:optimizing} for different ratios of $m/(k+s)$ between $\boldsymbol{y}$, and $\boldsymbol{g},\boldsymbol{z}$ in the absence of noise. We consider $\boldsymbol{g}^{(0)}$ and $\boldsymbol{z}^{(0)}$ as both random and solution vectors of \eqref{eq:solg} and \eqref{eq:initz}, respectively. The success rate and the number of iterations are averaged over 100 pulses. The results (Fig. \ref{fig:success}) show the effectiveness of BliPhaSu algorithm when the initializations $\boldsymbol{g}^{(0)}$ and $\boldsymbol{z}^{(0)}$ are set to the solutions of \eqref{eq:solg} and \eqref{eq:initz}, respectively.
	
	\subsection{Simple test with experimental data}
	\begin{figure}[t]
		\centering
		\includegraphics[width=1.0\linewidth]{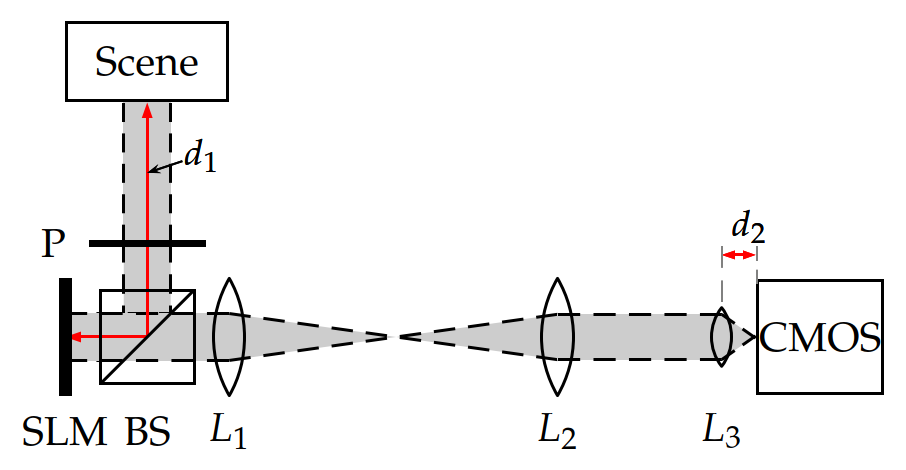}
		\caption{Experimental setup from \cite{rostami2021power} showing the polarizer (P), beamsplitter (BS), and spatial light modulator (SLM). The distance between the scene and the plane of the imaging system is $d_1$. The lenses $L_1$ and $L_2$ form the 4f-telescopic system projecting a wavefront from the SLM plane to the imaging lens $L_3$. The CMOS is a registering camera. The distance between this optical system and the sensor is $d_2$. }
		\label{fig:setup}
		\vspace{-1em}
	\end{figure}
	\begin{figure}[t]
		\centering
		\includegraphics[width=0.9\linewidth]{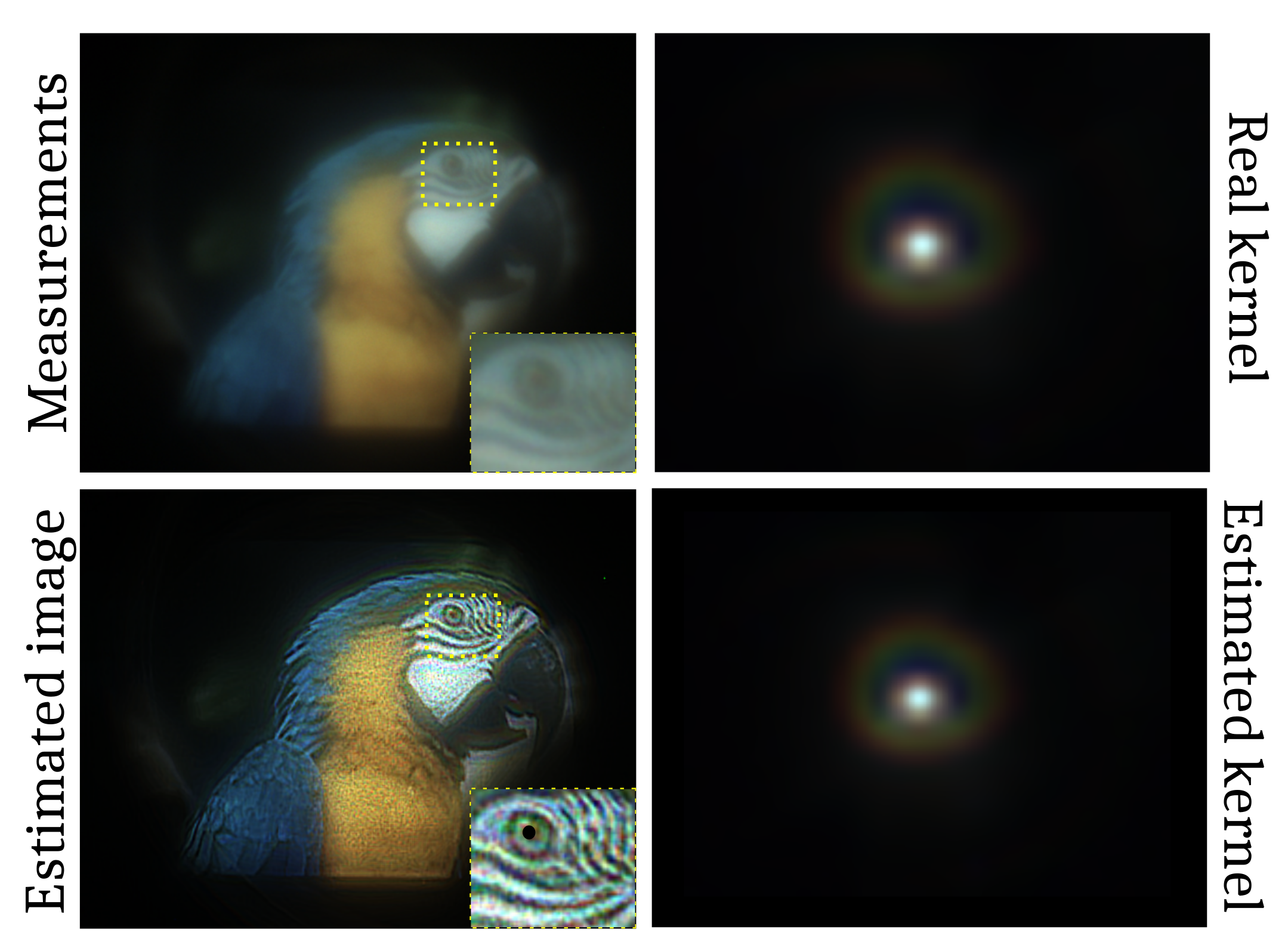}
		\caption{\footnotesize{Image reconstructed from experimental blurred observations (acquired by Igor Shevkunov at Tampere University) ('Measurements', top left) using BliPhaSu algorithm. The reconstructed image (bottom left) has a sub-pixel resolution of $0.86\mu m$. The true (top right) and recovered (bottom right) kernels are also shown.}}
		\label{fig:recons}
		\vspace{-1em}
	\end{figure}
	In this section we present a simple test employing experimental data to study the performance of the proposed algorithm. To that end, the optical setup depicted in Fig.~\ref{fig:setup} was built in \cite{rostami2021power} for intensity imaging that closely follows prior implementations on phase retrieval (coherent imaging) involving coded diffraction patterns \cite{katkovnik2017computational,bacca2019super}. We choose this intensity imaging scenario because the data measured by the sensor corresponds to the real-valued convolution between the signal and a kernel (point spread function) introduced by the system. Additionally, we can straightforward have knowledge about the matrices $\boldsymbol{B}$, $\boldsymbol{C}$ for $\boldsymbol{h}$, $\boldsymbol{x}$ respectively. For this very particular scenario, matrix $\boldsymbol{B}$ is a decimation matrix (resolution factor equal to four) since the kernel is assumed naturally sparse. And matrix $\boldsymbol{C}$ is the product between a decimation matrix (resolution factor equal to four) and the wavelet transform since the signal is assumed to be sparse in the wavelet domain. We would to stress that this test is an approximation to the problem of interest in \eqref{eq:basicproblem} since the difficulty to have access to experimental data and to guarantee the knowledge of matrices $\boldsymbol{B}$ and $\boldsymbol{C}$.
	
	In Fig.~\ref{fig:setup}, for the SLM a Holoeye phase-only GAEA-2-vis was employed, which has a spatial resolution of $4160\times2464$ with a pixel size of $3.74~\mu$m. The elements '$L_1$' and '$L_2$' models achromatic doublet lenses with diameter $12.7$~mm and focal distance of $50$~mm, BK7 glass lens '$L_3$' with diameter $6$~mm and focal distance $10$~mm. For the sensor the 'CMOS' Blackfly S board Level camera with the color pixel matrix Sony IMX264 is used with a pixel size of $3.45$ $\mu$m and total amount pixels of $2448\times2048$. The test 'scene' plane is displayed on a screen with LED illumination. To experimentally compare a reference kernel, the system in Fig. \ref{fig:setup} is calibrated using a fiber of diameter $200$ $\mu m$ as a point-source for white light in a dark room. With this optical system, we acquired blurred images ('Measurements') at the sensor through the SLM (Fig. \ref{fig:recons}). The same system is used to obtain a reference kernel. Then, the acquired experimental blurred data is used as input to BliPhaSu algorithm to increase the resolution of the target scene from $3.45$ $\mu m$ (sensor pitch size) to $0.86$ $\mu m$. These results show the effectiveness of our method to recover both image and kernel.
	
	\section{Summary}
	\label{sec:conclusion}
	We studied the blind deconvolution setting using low-resolution phaseless measurements. Our proposed non-convex optimization procedure accurately recovers both kernel and signal in the presence of noise. The BliPhaSu algorithm is shown to have linear convergence and better success rate using our initialization over random vectors. This is verified using experimental data from an actual optical measurement setup to show the recovery of both signal and kernel.
	
	\section*{Acknowledgments}
	The authors thank Igor Shevkunov, Vladimir Katkovnik, and Karen Egiazarian of the Computational Imaging Group at the Computing Sciences Unit, Faculty of Information Technology and Communication Sciences, Tampere University for their contributions in acquiring real blurred data to study the performance of the proposed algorithm. K. V. M. acknowledges support from the National Academies of Sciences, Engineering, and Medicine via Army Research Laboratory Harry Diamond Distinguished Postdoctoral Fellowship. S. P. acknowledges support from by the CIWIL project funded by ``Jane and Aatos Erkko'' and ``Technology Industries of Finland Centennial'' Foundations, Finland and EMET Research Institute, Colombia.
	
	\balance
	\bibliographystyle{IEEEtran}
	\bibliography{main}
	
\end{document}